\documentclass[11pt]{article}
\usepackage[utf8]{inputenc}
\usepackage{amsmath,amsthm,amssymb,fullpage,hyperref}
\newcommand{\ex}[2]{{\ifx&#1& \mathbb{E} \else \underset{#1}{\mathbb{E}} \fi \left[#2\right]}}
\newcommand{\pr}[2]{{\ifx&#1& \mathbb{P} \else \underset{#1}{\mathbb{P}} \fi \left[#2\right]}}
\newcommand{\var}[2]{{\ifx&#1& \mathsf{Var} \else \underset{#1}{\mathsf{Var}} \fi \left[#2\right]}}

\newcommand{\nope}[1]{}

\usepackage[style=alphabetic,backend=bibtex,maxbibnames=20,maxcitenames=6,giveninits=true,doi=false,url=true]{biblatex}
\newcommand*{\citet}[1]{\AtNextCite{\AtEachCitekey{\defcounter{maxnames}{2}}} \textcite{#1}}

\newcommand*{\citep}[1]{\cite{#1}}

\AtBeginBibliography{\small}%

\usepackage{enumitem}

\usepackage{titlesec}
\titlespacing*{\section}{0pt}{0.5\baselineskip}{0.25\baselineskip}
\titlespacing*{\subsection}{0pt}{0.4\baselineskip}{0.2\baselineskip}
 
\title{Multi-Central Differential Privacy}
\author{Thomas Steinke\thanks{IBM Research -- Almaden~\dotfill~\texttt{multicentral@thomas-steinke.net}}}
\date{\vspace{-6ex}}

\begin{document}
\maketitle

\begin{abstract}
Differential privacy is typically studied in the \emph{central model} where a trusted ``aggregator'' holds the sensitive data of all the individuals and is responsible for protecting their privacy. A popular alternative is the \emph{local model} in which the aggregator is untrusted and instead each individual is responsible for their own privacy. The decentralized privacy guarantee of the local model comes at a high price in statistical utility or computational complexity. Thus intermediate models such as the \emph{shuffled model} and \emph{pan privacy} have been studied in an attempt to attain the best of both worlds.

In this note, we propose an intermediate trust model for differential privacy, which we call the \emph{multi-central model}. Here there are multiple aggregators and we only assume that they do not collude nefariously. This model relaxes the trust requirements of the central model while avoiding the price of the local model. 
We motivate this model and provide some simple and efficient algorithms for it. We argue that this model is a promising direction for further research.
\end{abstract}

\section{Introduction}

Differential privacy \cite{DworkMNS06} is a formal standard for the protection of sensitive information about individuals in data analysis. Informally, differential privacy ensures that we cannot infer the attributes of any individual from the output of the algorithm. %
Differentially private algorithms randomly distort their output enough to obscure an individual's contribution -- but, hopefully, not too much, so that the output remains useful.

It is simplest to assume that the differentially private algorithm is executed by a trusted party -- the ``aggregator'' -- that has access to all the data and only the final output is released. This standard setup is referred to as the \emph{central model}. 

The central model requires individuals to trust the aggregator to securely store and process their data, but such a universally trusted entity may not exist. Thus it is important to also consider privacy threats arising from the computation process, rather than just the final output. 

There are a few approaches for avoiding the need for a trusted aggregator:
\begin{itemize}[topsep=0.5ex,itemsep=0.5ex,leftmargin=0.5cm]

\item \emph{Local randomization:} Certain algorithms, like randomized response \citep{Warner65}, perform the distortion needed to guarantee differential privacy at the individual level, before the data is aggregated. Since the individuals (or, rather, their personal devices) perform the distortion themselves, they need not trust any other party. This is what is known as the \emph{local model} of differential privacy \citep{KasiviswanathanLNRS11}. 

The minimal trust requirements and scalable algorithms of local differential privacy make it very attractive in practice. Google \cite{RAPPOR}, Apple \cite{AppleDP}, and Microsoft \cite{MicrosoftDP} have all deployed locally differentially private systems with millions of users.

Unfortunately, algorithms in the local model are necessarily less accurate than their counterparts in the central model \cite{BeimelNO08,KasiviswanathanLNRS11,ChanSS12b,DuchiJW18,Ullman18,BunNS18,DuchiR18,JosephMNR19,JosephMR20}. Essentially, each individual must randomly distort their own data as much as a trusted aggregator would distort the sum of everyone's data.
The variances of all these distortions add up, resulting in much larger error being necessary in the local model than in the central model.

\item \emph{Cryptography:} %
Various works \citep{DworkKMMN06,BeimelNO08,ShiCRCS11,NarayanH12,ChanSS12a,PettaiL15,BonawitzIKMMPRSS17,GoryczkaX17,Kairouz19} have employed variants of \emph{secure multiparty computation (SMC)} (a.k.a. \emph{secure function evaluation}) \citep{Yao82,Yao86,GoldreichMW87,LindellP08} to simulate having a trusted aggregator while only making trust assumptions akin to the local model. These tools typically provide a computational differential privacy guarantee \citep{MironovPRV09,McGregorMPRTV10} -- that is, we must make an assumption about the computational abilities of a potential adversary. %

However, these cryptographic protocols incur significant overheads. Building practical SMC systems remains a significant research challenge \citep{PinkasSSW09,HuangKE13,CarterMTB16}. At this stage, it is not feasible to scale SMC to thousands or millions of individuals who only have the computational and communication resources (and reliability) of a personal computer or smartphone.

\item \emph{Special assumptions:} Intermediate trust models between the central model and the local model have also been considered. These make some special-purpose assumption, with the goal of striking a better balance between realistic privacy and statistical utility.

The \emph{pan-private model} \cite{DworkNPR10} assumes that the aggregator is trustworthy now, but may not be trustworthy in the future. It requires the aggregator to only store distorted values, instead of the raw data, to protect against future intrusions. 

The \emph{shuffled model} \cite{CheuSUZZ19,ErlingssonFMRTT19} assumes that there are many individuals honestly participating in the protocol and that the aggregator receives their messages in an anonymous manner -- that is, the adversary does not know which message originated from which individual.

\end{itemize}
We propose another intermediate trust model between the central and local models and argue that it attains the best of both worlds -- a realistic trust model, accurate and practical algorithms, and, of course, a rigorous privacy guarantee.

\subsection{Our Model: Multi-Central Differential Privacy}

We now specify the setup of our model and the guarantee of multi-central differential privacy.

There are $m$ aggregators $A_1, \cdots, A_m$ and $n$ individuals with private data $x_1, \cdots, x_n$. The individuals will communicate with the aggregators and the aggregators will communicate with one another and, finally, produce some public output. 

Our trust assumption is that the aggregators do not collude nefariously. That is, we assume that at least one of the aggregators is trustworthy, but we do not know which one. 
More precisely, under multi-central differential privacy, the adversary chooses $m-1$ aggregators and $n-1$ individuals to control. The remaining individual to be protected and remaining trustworthy aggregator faithfully follow the protocol. The view of the adversary consists of all the information available to the aggregators it controls plus the final output of the trustworthy aggregator. It excludes only the internal information of the trustworthy aggregator and the individual to be protected. Differential privacy for the remaining individual's data must be provided with respect to this view. That is, the protocol is computationally $(\varepsilon,\delta)$-differentially private in the multi-central model if, for all trustworthy aggregators $j \in [m]$ and all individuals $i \in [n]$ to be protected and all datasets $x_1, \cdots, x_n, x_i'$ and all efficiently computable $S$, \[ \!\!\!\!\!\!\!\!\!\!\!\!\!\!\! \pr{}{\mathsf{View}_{A_1, \cdots, A_{j-1}, \mathsf{output}(A_j), A_{j+1}, \cdots, A_m}(x_1, \cdots, x_n) \in S} \!\le e^\varepsilon \cdot \pr{}{\mathsf{View}_{A_1, \cdots, \mathsf{output}(A_j), \cdots, A_m}(x_1, \cdots, x_i', \cdots, x_n) \in S} + \delta.\]

We remark that the number of aggregators $m$ can be viewed as a ``trust parameter.'' The larger $m$ is, the less reliant we are on trusting any individual aggregator. Setting $m=1$ simply recovers the usual central model, while $m=n$ is analogous to the local model. The statistical or computational cost of privacy also increases with $m$.

Note that this model does not guarantee utility in the presence of adversaries; utility may be more easily compromised by a single malicious party than privacy. 

\subsection{Motivation}

We envisage the $m$ aggregators as being a handful of large data centers and the $n$ individuals as millions of people with smartphones or personal computers and all communication happening via the internet using public-key cryptography. The final output can be freely shared.

We now suggest several scenarios in which our trust model is potentially realistic (or, at least, may be more realistic than the central model). 

\begin{itemize}[topsep=0.5ex,itemsep=0.5ex,leftmargin=0.5cm]
\item \emph{Insider/physical threats:} A major concern when individuals allow their data to be stored and processed in a data center is that an employee or someone who has gained physical access to the servers will obtain their sensitive data. 

The multi-central model protects against such insider threats, as long as no employee is granted access to systems in all of the data centers. Likewise, a physical attacker would need to break into all of the data centers to compromise the multi-central model.

\item \emph{Multiple jurisdictions:} Privacy laws are becoming increasingly complex and this is compounded by the fact that data often crosses international borders. For example, the European Union's General Data Protection Regulation (EU GDPR) restricts the transfer of private data abroad. %

The multi-central model may help address some of these international legal challenges. If a data center (and the relevant management) is placed in each applicable jurisdiction, then the data could be protected under the laws of all of these jurisdictions. Bypassing multi-central differential privacy and accessing the raw data in a legal fashion would presumably require adhering to the privacy laws of all jurisdictions. Users may have increased confidence in the privacy protections knowing that, in the event of a privacy breach, they potentially have legal recourse in multiple courts.

\item \emph{Trust as a Service:} A trusted organization can be contracted to serve as one of the aggregators. Thereby users need only trust that this organization follows the terms of its contract. In essence, this organization is providing a ``trust service''. Such a service would be useful for companies that are not themselves fully trusted by users or have incentives to breach privacy.

\item \emph{Collaborating Rivals:} The multi-central model provides a framework for companies that are usually fierce rivals to collaborate and, in the process, benefit the privacy of their users. In this framework, the companies would each take the role of one aggregator and their user bases would be combined for the purpose of data collection. Thereby the companies gain better statistical accuracy from the larger combined user base. Furthermore, users' privacy is protected by the rivalry between companies. That is, the multi-central model protects the user data, as long as we trust the companies to not collude to violate privacy.

\item \emph{Honest but curious:} The multi-central model protects against  overly curious aggregators -- since each aggregator only sees the data in an indecipherable form, they are not tempted to explore it. %

\end{itemize}
Needless to say, the multi-central model does not account for all possible threats. For example, it is likely that the aggregators will share software or a hardware provider; if this common software or hardware is compromised, then so too is the multi-central model. However, even local differential privacy is vulnerable to compromised software or hardware at the user end. Additionally, if all the aggregators belong to the same organization, then the leadership of that organization could presumably violate the multi-central privacy protections. This is why it is preferrable to have the aggregators belong to distinct organizations.

\subsection{Related Work}
We begin by noting that our proposed trust model is not novel. It has been widely studied, for example, in \emph{private information retrieval} \cite{ChorGKS95} and \emph{secret sharing} \cite{Shamir79,Blakley79}. Indeed, we seek to build upon these lines of work.

Furthermore, combining this trust model with differential privacy has been previously proposed \cite{Corrigan-GibbsB17}. However, to the best of our knowledge, this has not been seriously investigated. The purpose of this note is to begin such an investigation and to encourage further work.

The multi-central trust model is, of course, closely related to the central, local, and shuffled models. We briefly remark about how these compare. Clearly, any algorithm for the local model can be run in the multi-central model and any algorithm in the multi-central model can be run in the central model. This establishes the multi-central model as an intermediate trust model. 

Also, the shuffled model can be simulated in the multi-central model -- the aggregators must simply perform the shuffling of messages. Thus our model subsumes the functionality of the shuffled model.

Using secure multiparty computation, it is possible to simulate the central model in the multi-central model (with a computational differential privacy guarantee). Unfortunately, secure multiparty computation is not yet practical for general-purpose computations. Nonetheless, this establishes that there is no inherent ceiling to the potential of algorithms the multi-central model model. The only limitation to the multi-central model is our ability to devise and execute algorithms and the limits that would be present even in the central model.

\medskip

In the rest of this note, we explore some algorithms for the multi-central model. We believe these demonstrate that the multi-central model warrants further (empirical) exploration.

\section{Algorithms}

Having introduced and motivated the multi-central model, we now turn to demonstrating its utility by proposing several simple, accurate, and efficient private algorithms.

\subsection{Counting Queries}

Counting queries are a particularly simple and fundamental task. Namely, given $q : \mathcal{X} \to \{0,1\}$, the task is to privately approximate $\frac{1}{n} \sum_{i=1}^n q(x_i)$, where the private data of the individuals is $x_1, \cdots, x_n \in \mathcal{X}$. 
Our algorithm for this task is also simple and is based on a \emph{linear secret sharing scheme} \cite{Shamir79,Blakley79}.

Choose a modulus $p > n$. Each individual $i \in [n]$ generates independent uniformly random values $v_i^1, \cdots, v_i^{m-1} \in \mathbb{Z}_p$ and sets $v_i^m = q(x_i) - \sum_{j=1}^{m-1} v_i^j \mod{p}$. Then $v_i^j$ is transmitted to $A_j$ (using public-key cryptography) for each $j \in [m]$. 

If the aggregators where to combine all their values, they can recover $q(x_i) = \sum_{j=1}^m v_i^j$. However, any $m-1$ aggregators see only uniformly random values and have no information about $q(x_i)$. This is the security property of the secret sharing scheme and the basis of the multi-central differential privacy guarantee.

In the protocol, each aggregator $j \in [m]$ computes $v^j=\sum_{i=1}^n v_i^j$ and outputs a noisy sample $V^j \leftarrow \mathcal{N}(v^j, \sigma^2)$ from the continuous or discrete Gaussian \cite{CanonneKS20}. The addition of noise ensures multi-central $(\rho=1/2\sigma^2)$-concentrated differential privacy \cite{BunS16}, which entails  multi-central $(\varepsilon=1/2\sigma^2 + \sqrt{2\log(1/\delta)/\sigma^2},\delta)$-differential privacy \cite{DworkKMMN06} for all $\delta>0$.

The final output is $\frac{\sum_{j=1}^m V^j \mod{p}}{n}$, which is distributed as $\mathcal{N}\left( \frac{1}{n} \sum_{i=1}^n q(x_i) , \frac{m \cdot \sigma^2}{n^2} \right) \mod{\frac{p}{n}}$. Assuming $p \gg n$ and $m \cdot \sigma^2 \ll n^2$, this value will be close to the desired value. The accuracy guarantee degrades as $m$ gets larger. We note that $m=1$ recovers the accuracy guarantee obtainable in the central model and $m=n$ recovers the accuracy guarantee obtainable in the local model. 

This algorithm is also very efficient. Each individual must transmit $O(\log p) = O(\log n)$ bits per aggregator in one round. The main computational task is encrypting the messages for transmission. The aggregators must simply decrypt the values and add them up along with some noise.

\subsection{Heavy Hitters}

Identifying heavy hitters is a well-studied and important task in differential privacy, particularly in the local model \cite{MishraS06,HsuKR12,RAPPOR,BassilyS15,AppleDP,QinYYKXR16,BassilyST17,BunNS18}. The key task is to be able to privately estimate the frequency within the dataset of any element. (Sophisticated reductions on top of such a frequency estimator can then be used to find the frequent elements.)

We can efficiently construct a frequency estimator for multi-central differential privacy using the previous algorithm for statistical queries combined with linear sketching, such as the Johnson-Lindenstrauss transform. In particular, our algorithm produces a sketch subject to multi-central $(\rho = \frac12 \varepsilon^2)$-concentrated differential privacy \cite{BunS16} which can be used to estimate the frequency of any element with additive relative error $\alpha=O\left(\frac{\sqrt{m\log(1/\beta)}}{\varepsilon n} + \sqrt{\frac{\log(1/\beta)}{\ell}}\right)$, where $\beta$ is the allowed failure probability and $\ell$ is the size of the sketch. Each individual must send $O(\ell \cdot \log n)$ bits to each aggregator. (The algorithm also requires $O(\log|\mathcal{X}| \cdot \log(1/\beta))$ bits of shared randomness for the sketch \cite[Thm.~2.2]{ClarksonW09}, where $\mathcal{X}$ is the data universe.) Multiple queries can be answered using the same sketch.

For $m=O(1)$, this provides utility comparable to the central model.

\subsection{Computationally-Secure Secret Sharing \& Decision Trees} 
The previous two algorithms use information-theoretic secret sharing. However, we can reduce the communication in some settings by using computational \emph{function secret sharing}.
\citet{BoyleGI15,BoyleGI16} (see also \cite{WangYGVZ17,Wang17}) show how to secret share any family of queries represented by a decision tree of size $\ell$ with secret shares of size $O(\lambda\cdot\ell )$, where $\lambda$ is the security parameter (i.e., key length). These constructions are very efficient and only use symmetric-key cryptography (specifically, pseudorandom generators), rather than more exotic cryptographic primitives.

This yields an efficient algorithm in the multi-central model for threshold queries \cite{DworkNPR10,BunNSV15} over domain $\mathcal{X} \subset \mathbb{R}$, which can be computed by a decision tree of size $\ell=O(\log|\mathcal{X}|)$. This also yields an alternative frequency estimation algorithm for heavy hitters in which the share size and, hence, communication per aggregator is $O(\lambda\cdot\min\{\log|\mathcal{X}|,\log(1/\alpha\beta)\})$ bits, rather than $O(\log(n)\cdot\log(1/\beta)/\alpha^2)$ as above. This is a significant improvement in the high accuracy (small $\alpha$) regime.

More generally, function secret sharing can be used to significantly reduce communication in the multi-central model. Answers to a structured family of queries (e.g., thresholds) can be compressed. However, compression is possible even in unstructured settings: For example, suppose we need to answer $k$ arbitrary counting queries. If we use the information-theoretic secret sharing algorithm above $k$ times, then the communication will grow linearly in $k$, which could be prohibitive. However, it is possible to instead have each individual pick one out of the $k$ queries at random and only answer that one. Function secret sharing allows the individual to transmit this single answer to the aggregators without the aggregators learning which query was answered and, crucially, the communication grows only logarithmically in $k$. Effectively, the individual sends a vector of answers where at most one entry is nonzero and this can be compressed. This approach introduces some statistical noise due to sampling, but, depending on the parameters, should still provide sufficiently accurate answers. In terms of privacy, each query answer should be privatized as before, but, since the adversary does not know which query the individual chose to answer, their privacy is amplified \cite{CheuSUZZ19,ErlingssonFMRTT19} and the final guarantee is comparable to that of the less efficient algorithm.

\subsection{Secure Multiparty Computation \& Selection}
A general-purpose tool for the multi-central model is secure multiparty computation (SMC). The individuals can use a simple secret sharing scheme to send their data to the aggregators, who can then use SMC to compute the desired differentially private output as if they were a single trusted aggregator. This protocol is very efficient for the individuals and allows a rich set of analyses to be run, but may be computationally expensive for the aggregators (although not as expensive as including all individuals in the SMC protocol). 

The practicality of the SMC will depend on the task being run. One important task is selection -- given a dataset $x$ and a set of (counting) queries $q_1, q_2, \cdots, q_k$, we wish to privately output some approximation to $\underset{\ell \in [k]}{\mathsf{argmax}} ~ q_\ell(x)$. The classic algorithm for selection is the exponential mechanism \cite{McSherryT07}. An alternative is the report-noisy-max algorithm \cite{DworkR14}, which outputs $\underset{\ell \in [k]}{\mathsf{argmax}} ~ q_\ell(x) + \xi_\ell$, where $\xi_1, \cdots, \xi_k$ are independent Laplace noise scaled to the global sensitivity of the queries. We can implement a variant of report-noisy-max in the multi-central model using fairly limited SMC as follows.

First, every individual $i \in [n]$ computes secret shares $v_{i,j,\ell} \in \mathbb{Z}_p$ such that $\sum_{j \in [m]} v_{i,j,\ell} = q_\ell(x_i) \mod{p}$ for all $\ell \in [k]$. Each $v_{i,j,\ell}$ is then sent to aggregator $j$. Each aggregator $j \in [m]$ computes $\hat{v}_{j,\ell} := \sum_{i \in [n]} v_{i,j,\ell} + \xi_{j,\ell} \mod{p}$, where $\xi_{j,\ell}$ is (discrete) Laplace noise \cite{CanonneKS20}. Finally, SMC is used to compute \[\ell^* := \underset{\ell \in [k]}{\mathsf{argmax}} ~ \left( \sum_{j \in [m]} \hat{v}_{j,\ell} \mod{p} \right)\] and the value $\ell^*$ is released. This is differentially private as long as the SMC only reveals the argmax and the adversary does not know the noise added by at least one of the aggregators. The utility is also comparable to that of the exponential mechanism, as long as the number of aggregators $m$ is small. In terms of computation, the SMC must only compute the modular sum and the argmax. There is hope that this special-purpose task can be performed practically, but further investigation is needed.

We remark that efficient selection is believed to \emph{not} be possible in the shuffled model \cite{CheuSUZZ19,GhaziGKPV19}. Thus this suggests a possible advantage of the multi-central model. Selection is an important tool, and is a critical subroutine in more complex systems \cite{HardtR10}. Implementing selection will allow other things to be built on top of it.

\section{Conclusion and Further Work}

This note motivates the study of the multi-central model of differential privacy and proposes several basic algorithms for the model. We hope that this motivates and serves as a basis for further exploration, including experimental evaluations.

We hope to develop more algorithms, guarantee accuracy in the presence of faults (using tools like \emph{verifiable secret sharing} \citep{ChorGMA85,Corrigan-GibbsB17}), and also explore other trust models. %

\printbibliography

\end{document}